\documentclass[preprintnumbers,showpacs,amsmath,amssymb,floatfix,9pt,prd,onecolumn,
superscriptaddress,nofootinbib]{revtex4}
\usepackage{latexsym}
\usepackage{epsfig}
\usepackage{amssymb}

\begin{document}

\title{Charged star in (2+1)-dimensional gravity}

\author{Ayan Banerjee, Farook Rahaman}
\email{ayan_7575@yahoo.co.in, rahaman@iucaa.ernet.in} \affiliation{Department of
Mathematics, Jadavpur University, Kolkata, India-700032}
\author{Tanuka Chattopadhyay}
\email{tanuka@iucaa.ernet.in} \affiliation{Department of Applied Mathematics, Calcutta University, Kolkata, India
92 A.P.C. Road, Kolkata -700009}

\begin{abstract}
We obtain a new class of exact solutions for the Einstein-Maxwell system in static spherically
symmetric charged star in (2+1)-dimensional gravity. In order
to obtain the analytical solutions we treat the matter distribution anisotropic in nature admitting
linear or nonlinear equation of state and the electric field intensity was specified.
By choosing a suitable choice of mass function m(r), it is possible to integrate the
system in closed form. All the solution, which are obtained in both linear and nonlinear
cases are regular at the center and well behaved in the stellar interior.

\end{abstract}

\maketitle

\section{Introduction}
 For finding the comprehensive picture of today's universe, researchers
 are trying to solve  Einstein's equation with different inputs for
 physical contact (dust, perfect fluid, etc.) and feature (charge, rotation, anisotropic matter,
 etc.). In relativistic astrophysics, the study of static anisotropic
 fluid spheres attracts the interest of researchers after the extensive investigation by
 Bowers and Liagn \cite {Bowers}, and the theoretical investigations by R. Ruderman \cite{Ruderman},
 showing that the nuclear matter may be
 anisotropic (i.e. radial pressure may not be equal to the tangential pressure)
  in a very high density range ($\rho > 10^{15}$ $g/cm^3$)
 where nuclear interaction need to be treated relativistically. A large
 number of study have been found corresponding to the anisotropic
 matter applied to the relativistic star when the gravitational
 field is taken to be static spherically symmetric which is commonly
 done by matching two given and known space-time at the junction surface.

   In recent year, lower dimensional gravity has turn into a
special attention for its simplicity to describe the geometry of
space-time. Specially it enlighten some ambiguity which appeared in
four dimensional gravity. For example, the study of black hole
become more complicated in four dimensional gravity. To
overcome these difficulties and get reasonable acceptable results,
researchers have resorted to the models in the lower dimensional
gravity. On this respect most revolutionary work had been done by
Ba\~{n}ados, Teitelboim and Zanelli \cite{BTZ}, with the discovery of
(2+1) dimensional black hole with negative cosmological constant
known as BTZ black hole which has many similarity with the
characteristic of four dimensional black hole. The interesting
features of BTZ black holes are, it is a solution of low energy
string theory with a non-vanishing antisymmetric tensor and it can
be matched to the exterior of a (2+1) dimensional perfect fluid
star. With this point of view Cruz and Zanelli \cite{Cruz},
 studied the stability of these stars and puts an upper
limit for mass considering generic equation of state
$P = P(\rho)$. Garc\'{i}ýa and Campuzano \cite{Garc}, have derived all static
circularly symmetric perfect fluid 2+1 dimensional gravity and
made an important conclusion that by a simple dimensional reduction
one can get a  2+1 perfect fluid solution with constant energy
density that can be derived from the Schwarzschild interior metric
by a comparison with (2+1) and (3+1) gravity. Mann and Ross \cite{Mann},
shown that a (2+1) dimensional star filled with dust
(P=0), might collapsed to a black hole under certain conditions.
David Garfinkel \cite{Garfinkle}, have found an exact solution
for the critical collapse of a scalar field in closed form
for (2+1) dimensional gravity with negative cosmological
constant. Another interesting solution have found by Paulo M. S\'{a} \cite{Paulo}
for three-dimensional perfect fluid stars with polytropic
equation of state of the form $P = K\rho^{1+\frac{1}{n}}$, where
`n' is polytropic index and 'K' is polytropic constant. Recently,
Sharma \emph{et al} \cite{Sharma} have found a star solution with
simple analytical form by choosing the form of the mass
function m(r). Motivated by the work we have found
a new star solution, assuming the interior space-time
is Finch-skea type for (2+1) dimensions and match the
exterior with anti-de Sitter BTZ metric in \cite{Ayan}.

 Our objective is to find a new solutions of the Einstein-Maxwell system satisfying
the necessary criteria of a neutral star solution in (2+1)-dimensional gravity.
A spherically symmetric charged ideal fluid solution of Einstein field
equation has been studied by N. $\ddot{O}$zdemir \cite{Ozdemir} in the presence of the
cosmological constant. Mak and Harko \cite{M.K.Mak} obtained an exact analytical solution
for describing the interior of a charged strange quark star under the assumption of spherical
symmetry and the existence of a one-parameter group of conformal motions. A well studied
of charged objects in the frame of General Relativity, by solving the coupled Einstein-Maxwell
field equations has been found in \cite{ Komathiraj,R.Sharma,Patel,L.K. Patel,Thirukkanesh}.
Varela \emph{et al} \cite{Varela} have found an analytical solutions for self-gravitating, charged,
anisotropic fluids with linear or non-linear equations of state and obtain more flexibility
in solving the Einstein-Maxwell equations.

 In this paper, we discuss a new type of charged star in (2+1) dimensional
 gravity assuming the matter distribution is anisotropic in nature
 admitting a linear or nonlinear EOS. We match our interior solutions corresponding
 to the exterior charged BTZ metric with zero spin. The solution which we obtained, satisfied
 all the regularity conditions for a star solution and its simple analytic form helps
 to evaluate the physical parameters in a simple manner for the star solution.

\section{ Field equations and Interior geometry}
 We assume that the gravitational field should be static and spherically symmetric for describing the
internal geometry of a charged relativistic compact object. For describing
such a configuration in (2+1)-dimensional gravity the line element is given by
 \begin{equation}
 ds^{2} = -e^{\nu(r)}dt^{2}+ e^{\mu(r)}dr^{2} + r^{2} d\theta^{2},\label{eq1}
 \end{equation}
where $\nu(r)$ and $\mu(r)$ are the two unknown metric functions.
Following an earlier treatment \cite{Rahaman}, the Einstein-Maxwell field equations for
anisotropic fluid in the presence of negative cosmological constant ($\Lambda ~< 0$),
for the space-time described by the metric Eq.~(\ref{eq1}), yield (we set G =C= 1)
\begin{eqnarray}
8\pi\rho +E^2+\Lambda &= &\frac{\mu^{\prime}e^{-\mu}}{2r},\label{eq2}\\
8\pi p_r-E^2 -\Lambda &= & \frac{\nu^{\prime}e^{-\mu}}{2r},\label{eq3}\\
 8\pi p_t+E^2 -\Lambda &= & \frac{e^{-\mu}}{2}\left(\frac{1}{2}{\nu^{\prime 2}}+{\nu^{\prime\prime}}-\frac{1}{2}\nu^{\prime}\mu^{\prime}\right),\label{eq4}\\
 \sigma (r) &= & \frac{e^{-\frac{\mu}{2}}}{4\pi r}\left(rE\right)^{\prime},\label{eq5}
\end{eqnarray}
where prime denotes the derivative w.r.t the radial parameter r. The
Eq.~(\ref{eq5}), can equivalently be expressed in the form
\begin{equation}
E(r)= \frac{4\pi}{r}\int^{r}_0 x  \sigma(x)e^{\frac{\mu(x)}{2}}dx~=~\frac{q(r)}{r},\label{eq6}
 \end{equation}
where, q(r) is total charge of the sphere under consideration and
$\rho$, $p_r$ and $p_t$ represents the energy density, radial pressure and
transverse pressure, respectively. The quantities associated
with the electric field are the electric field intensity (E) and the volume
charge density ($\sigma$), and  Eqs.~(\ref{eq2})-(\ref{eq5}), are invariant under the transformation
$E\rightarrow -E$ and $\sigma \rightarrow -\sigma$. Now, due to the energy density
of the matter and the electric energy $(\frac{E^2}{8\pi})$ density, the mass of the star within
a spherical shell of radius r is takes of the form
\begin{equation}
m(r)= \int^{r}_0 2\pi  ~\widetilde{r}\left(\rho+\frac{E^2}{8\pi}\right)~d\widetilde{r}.\label{eq7}
 \end{equation}
Now, by integrating the
Eq.~(\ref{eq2}), we obtained
\begin{equation}
8m(r)=C-e^{-\mu(r)}-\Lambda r^2,\label{eq8}
 \end{equation}
where C is integrating constant.
Corresponding to an earlier work \cite{Sharma}, we set $C=1$ and assume $\mu(r)= Ar^2$ so as to
 ensure regular behaviour of the mass function at the centre i.e., $m(r)=0$ at r=0. The mass function is then obtained as
\begin{equation}
8m(r)=1-e^{-Ar^2}-\Lambda r^2,\label{eq9}
 \end{equation}
From Eq.~(\ref{eq9}), using Eq.~(\ref{eq7}) we obtain
\begin{equation}
2\pi\left(\rho+\frac{E^2}{8\pi}\right)=-\frac{\Lambda}{4}+\frac{A}{4}e^{-Ar^2}.\label{eq10}
\end{equation}
Now, it is necessary to evaluate the energy density $\rho$ from Eq.~(\ref{eq10}), so we
specify the electric field intensity E, though the various choices for E is possible but only
a few are physically acceptable features in the stellar interior. For our purpose we set
\begin{equation}
E^2= Be^{-Ar^2}-\Lambda,\label{eq11}
 \end{equation}
where $B < 0$ is constant parameter and $\Lambda (< 0)$ is cosmological constant.
The electric field given in Eq.~(\ref{eq11}) vanishes at the centre of the star i.e., E(0) = 0 when
 $\Lambda$ =B and remains continuous and positive at the interior of
star for relevant choices of the constants parameters. With
the choice Eq.~(\ref{eq11}) we can obtain the energy density $\rho$ from Eq.~(\ref{eq10}), which
 is given by
\begin{equation}
\rho=\frac{1}{2\pi}\left[\frac{A-B}{4} e^{-Ar^2}\right],\label{eq12}
 \end{equation}
which gives the central density as
\begin{equation}
\rho(r=0)=\frac{1}{2\pi}\left[\frac{A-B}{4}\right].\label{eq13}
 \end{equation}
The energy density and electric field intensity are plotted in Fig. 1. Note that to determine
the unknown metric potential $\nu(r)$, we select an EOS corresponding to the material
composition of the star,  given by
\begin{equation}
p_r = p_r\left(\rho,\alpha_1,\alpha_2\right),\label{eq14}
\end{equation}
where $\alpha_1$ and $\alpha_2$ are two  arbitrary constants and the radius $R$ of the star can be obtained by ensuring that
\begin{equation}
p_r\left(\rho(R),\alpha_1,\alpha_2\right) = 0.\label{eq15}
\end{equation}
We assume that the  EOS corresponding to (\ref{eq14}), in the
rest frame of anisotropic fluid distribution, the energy density $\rho$ and the pressures p may by related
with linear or non-linear EOS in nature and accordingly we consider the both possibilities separately.
\begin{figure*}[thbp]
\begin{tabular}{rl}
\includegraphics[width=9cm]{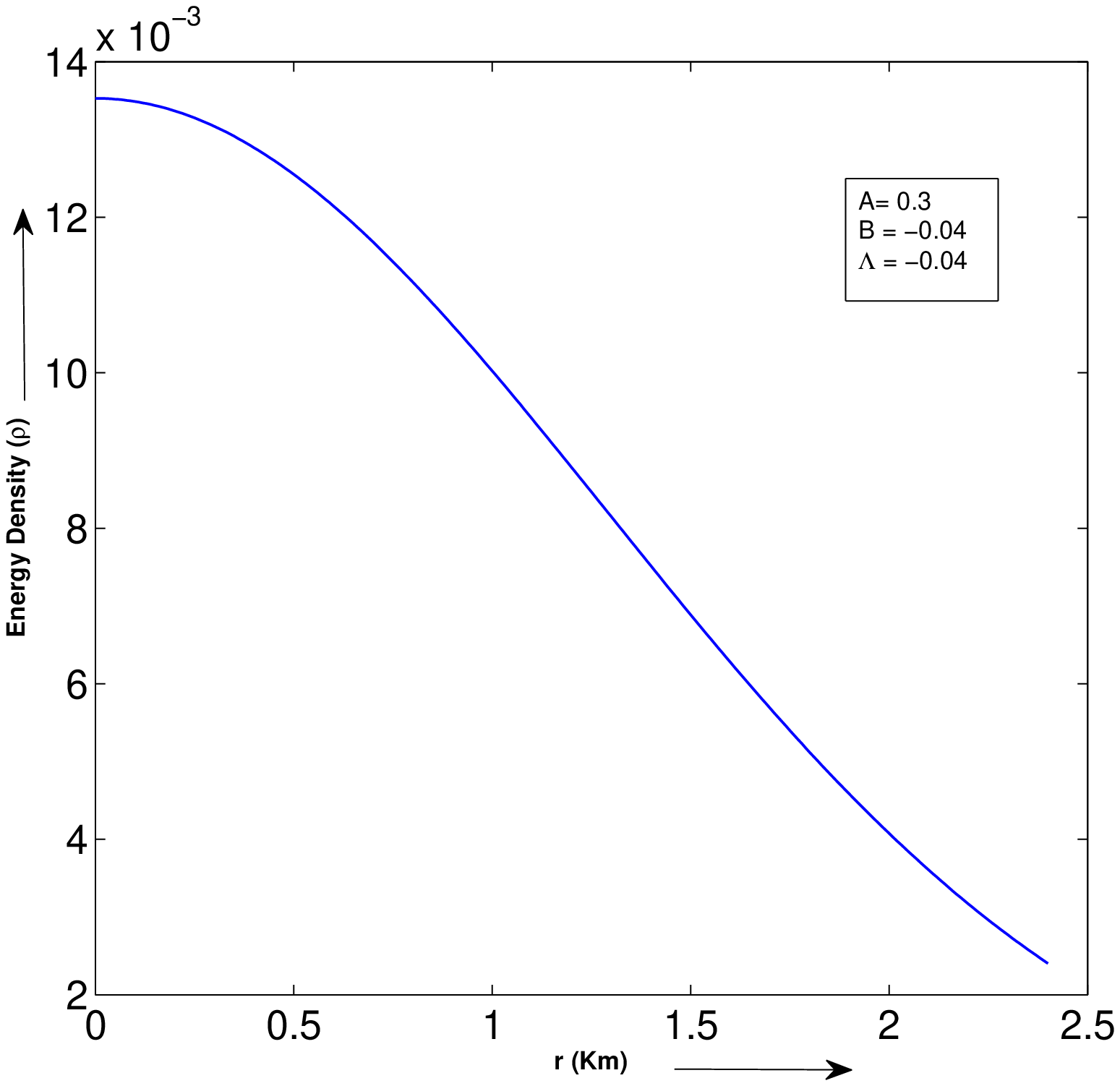}&
\includegraphics[width=9cm]{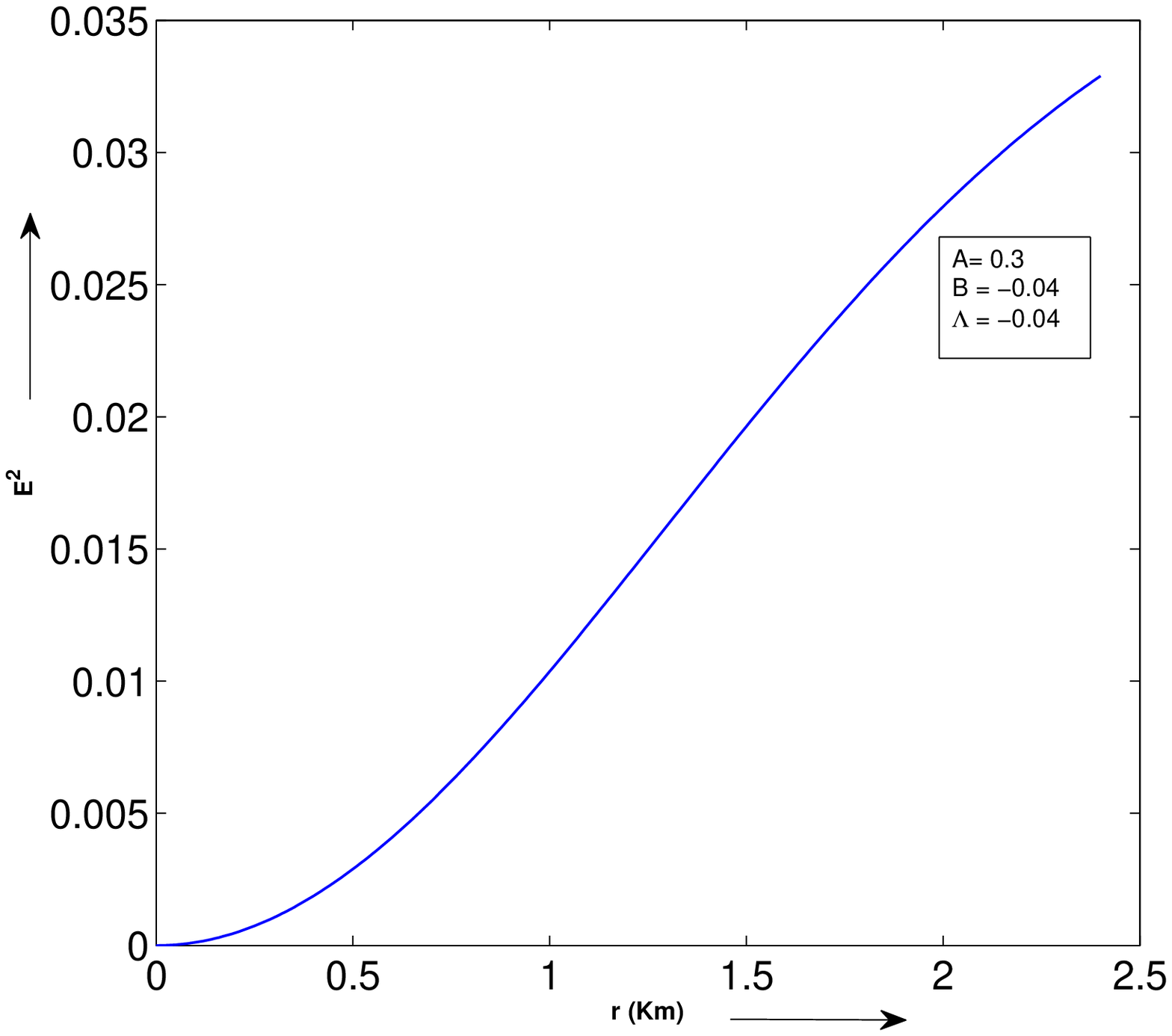} \\
\end{tabular}
\caption{ Variation of the energy-density `$\rho$' (left)  and the electric field intensity, `$E^2$' (right) at the stellar interior.}
\end{figure*}

\subsection{Solutions admitting a linear EOS}
Firstly we consider the linear EOS:
\begin{equation}
p_r=\alpha_1 \rho+\alpha_2,\label{eq16}
 \end{equation}
where $\alpha_1$ and $\alpha_2$ are constant parameters.
For the choice of (\ref{eq16}), the system of Eqs. (\ref{eq2})-(\ref{eq5}) can be solved analytically and we get
\begin{eqnarray}
\nu(r) &=& C_1+\left[\alpha_1 \left({A-B}\right)-B\right]r^2 +\frac{8\pi\alpha_2}{A}e^{Ar^2},\label{eq17}\\
p_r(r) &=& \frac{\alpha_1}{8\pi} \left({A-B}\right)e^{-Ar^2} +\alpha_2,\label{eq18}\\
p_t(r) &=& \frac{1}{8\pi} \left[\left(\alpha_1(A-B)-B\right)\left(\alpha_1(A-B)-(A+B)\right)r^2e^{-Ar^2}
+\left(\alpha_1(A-B)-B\right)(e^{-Ar^2}+16\pi\alpha_2r^2)\right. \nonumber\\
&&\left.  +64\pi^2\alpha_2^2 r^2 e^{Ar^2}+8\pi\alpha_2(1+Ar^2)-  Be^{-Ar^2}+2\Lambda \right].\label{eq19}
\end{eqnarray}
where $C_1$ is integrating constant. The measure of anisotropy is given by
\begin{eqnarray}
 \Delta  = p_t-p_r &=& \frac{1}{8\pi} \left[\left(\alpha_1(A-B)-B\right)\left(\alpha_1(A-B)-(A+B)\right)r^2e^{-Ar^2}
+\left(A-B\right)\left(1+2\alpha_1\right)8\pi\alpha_2r^2 \right. \nonumber\\
&&\left.  +64\pi^2\alpha_2^2 r^2 e^{Ar^2}- 2Be^{-Ar^2}+2\Lambda \right].\label{eq20}
\end{eqnarray}
It is a measure of the anisotropic  pressure of the fluid comprising the charged star.
This force due to the anisotropic nature is directed outward when $p_t > p_r$ $\Leftrightarrow $ $\Delta  > 0$
and inward if $p_t < p_r$ $\Leftrightarrow $ $\Delta  < 0$. $p_t = p_r$ i.e., $\Delta  = 0$ corresponds to the particular case
of an isotropic pressure of the charged fluid star.

\subsection{Solutions admitting a non-linear EOS}
Secondly, we consider the non-linear EOS:

\begin{equation}
p_r=\beta_1 \rho+\frac{\beta_2}{\rho},\label{eq21}
 \end{equation}
where $\beta_1$ and $\beta_2$ are constant parameters
and solving the system of equations (\ref{eq2})-(\ref{eq5}) analytically we get
\begin{eqnarray}
\nu(r) &=& C_2+\left[\beta_1 \left({A-B}\right)-B\right]r^2 +\frac{\beta_2}{A}\left(\frac{32\pi^2}{A-B}\right)e^{2Ar^2},\label{eq22}\\
p_r(r) &=& \beta_1 \left(\frac{A-B}{8\pi}\right)e^{-Ar^2} +\beta_2\left(\frac{8\pi}{A-B}\right)e^{Ar^2},\label{eq23}\\
p_t(r) &=& \frac{1}{8\pi} \left[\left(\beta_1(A-B)-B\right)^{2}r^2e^{-Ar^2}
  +\left(\beta_1(A-B)-B\right)(1-Ar^2)e^{-Ar^2}+\frac{\beta_2}{2}\left(\frac{128\pi^2}{A-B}\right)(1+3Ar^2)e^{Ar^2}\right. \nonumber\\
&&\left.  + \frac{\beta_2}{4}\left(\frac{128\pi^2}{A-B}\right)r^2e^{Ar^2}
        \left(\beta_2 \left(\frac{128\pi^2}{A-B}\right)e^{2Ar^2}+4\left(\beta_1(A-B)-B\right)\right)
        -Be^{-Ar^2}+2\Lambda \right].\label{eq24}
\end{eqnarray}
where $C_2$ is also integrating constant and the measure of anisotropy is given by

\begin{eqnarray}
 \Delta  = p_t-p_r &=& \frac{1}{8\pi} \left[\left(\beta_1(A-B)-B\right)^{2}r^2e^{-Ar^2}
  -\left(\beta_1(A-B)-B\right)Ar^2 e^{-Ar^2}+3\beta_2 A\left(\frac{64\pi^2}{A-B}\right)r^2e^{Ar^2}\right. \nonumber\\
&&\left.  + \frac{\beta_2}{4}\left(\frac{128\pi^2}{A-B}\right)r^2e^{Ar^2}
        \left(\beta_2 \left(\frac{128\pi^2}{A-B}\right)e^{2Ar^2}+4\left(\beta_1(A-B)-B\right)\right)
        -2 Be^{-Ar^2}+2\Lambda  \right],\label{eq25}
\end{eqnarray}

The measure of anisotropy vanishes i.e., $\Delta  = 0$ corresponds to the particular case
which depend on the constant parameters.

\section{Junction conditions}
The  electrovaccum exterior space-time of the circularly symmetric static
star is described by the charged BTZ black hole \cite{BTZ,ccj} can be written in the following form as
\begin{equation}
ds^{2} = -\left(-M_0 - \Lambda{r^2}-Q^2\ln r\right)dt^2 + \left(-M_0 -
\Lambda{r^2}-Q^2\ln r \right)^{-1}dr^2 + r^2d\theta^2,\label{eq26}
\end{equation}
where, the parameter $M_0$ is the conserved mass associated with asymptotic invariance
under time displacements and the parameter Q is total charge of the black
hole. At the boundary  $r=R$, continuity of the metric potentials yield the following junction conditions:
\begin{eqnarray}
e^{\nu(R)} &=& -M_0-\Lambda R^2-Q^2\ln R,\label{eq27}\\
e^{-\mu(R)} &=& -M_0-\Lambda R^2 -Q^2\ln R.\label{eq28}
\end{eqnarray}
Across the boundary r = R of the star together with the condition that the radial
pressure must vanish at the surface i.e., $p_r(r = R) = 0$, help us to determine these
constants. Now, using the metric function $\mu(r)= AR^2$, in Eq. (28), we obtain the following
solutions

\subsection{Solution for linear EOS }
the constant $C_1$ and the radius of the star obtained from Eqs.(\ref{eq17})-(\ref{eq18})
\begin{eqnarray}
A &=& -\frac{1}{R^2}\ln \left(-M_0-\Lambda R^2-Q^2\ln R\right),\label{eq29}\\
C_1 &=& \ln \left( -M_0-\Lambda r^2-Q^2\ln R\right)-(\alpha_1(A-B)-B)R^2-\frac{8\pi\alpha_2}{A\left( -M_0-\Lambda r^2-Q^2\ln R\right)},\label{eq30}\\
R &=& \frac{1}{\sqrt{A}}\left[\ln\left(-\frac{\alpha_1}{\alpha_2}\left(\frac{A-B}{8\pi}\right)\right)\right]^{\frac{1}{2}}.\label{eq31}
\end{eqnarray}

\subsection{Solution for non-linear EOS }
here also, the constant $C_2$ and the radius of the star obtained from Eqs.(\ref{eq22})-(\ref{eq23})
\begin{eqnarray}
A &=& -\frac{1}{R^2}\ln \left(-M_0-\Lambda R^2-Q^2\ln R\right),\label{eq32}\\
C_2 &=& \ln \left( -M_0-\Lambda r^2-Q^2\ln R\right)-(\beta_1(A-B)-B)R^2-\frac{\beta_2}{A}\left(\frac{32\pi^2}{A-B}\right)e^{2AR^2},\label{eq33}\\
R &=& \frac{1}{\sqrt{2A}}\left[\ln \left(-\frac{\beta_1}{\beta_2}\left(\frac{A-B}{8\pi}\right)^{2}\right)\right]^{\frac{1}{2}}.\label{eq34}
\end{eqnarray}

\section{Physical Analysis}

Now, we are in a position to investigate the physically meaningful interior solution for static fluid spheres.
In this case Einstein's gravitational field equations must satisfy certain physical requirements. 
Some of the physically acceptable conditions have been generally recognized to be crucial for anisotropic matter distribution \cite{Harrera,Mak} as:

 $\bullet$ Inside the star $r < R$, the density $\rho$ and the pressure p are everywhere positive and finite.

$\bullet$ The space-time is assumed  not to possess an event horizon i.e., $2m(r) < r$.

$\bullet$ The radial and tangential pressures should be decreasing functions of r.

$\bullet$ To keep the centre of the space-time regular, we enforce both $\rho^{\prime}_r(0) = p^{\prime}_ r(0) = 0$ and $p_r(0) = p_t(0)$.

$\bullet$ The gradients $\frac{d\rho}{dr}$ , $\frac{dp_r}{dr}$ and $\frac{dp_t}{dr}$ should be negative.

$\bullet$ The radial pressure $p_r$ must vanish but the tangential pressure $p_t$ may not vanish at the boundary
r = R of the star.

$\bullet$ Radial and transverse sound velocity should be less than unity i.e.  $0 \leq v_{sr}^2 (=\frac{dp_r}{d\rho}) \leq 1$ and $ 0 \leq v_{st}^2 (=\frac{dp_t}{d\rho}) \leq 1$.

The parameters can easily be adjusted to produce models where the speed of sound
is much less than unity. The behavior of the model
is illustrated best in terms of graphs of the matter variables and the gravitational potentials.
Here, we are trying to discuss all the restriction for both cases step by step
as given below:

\begin{figure*}[thbp]
\begin{tabular}{rl}
\includegraphics[width=9cm]{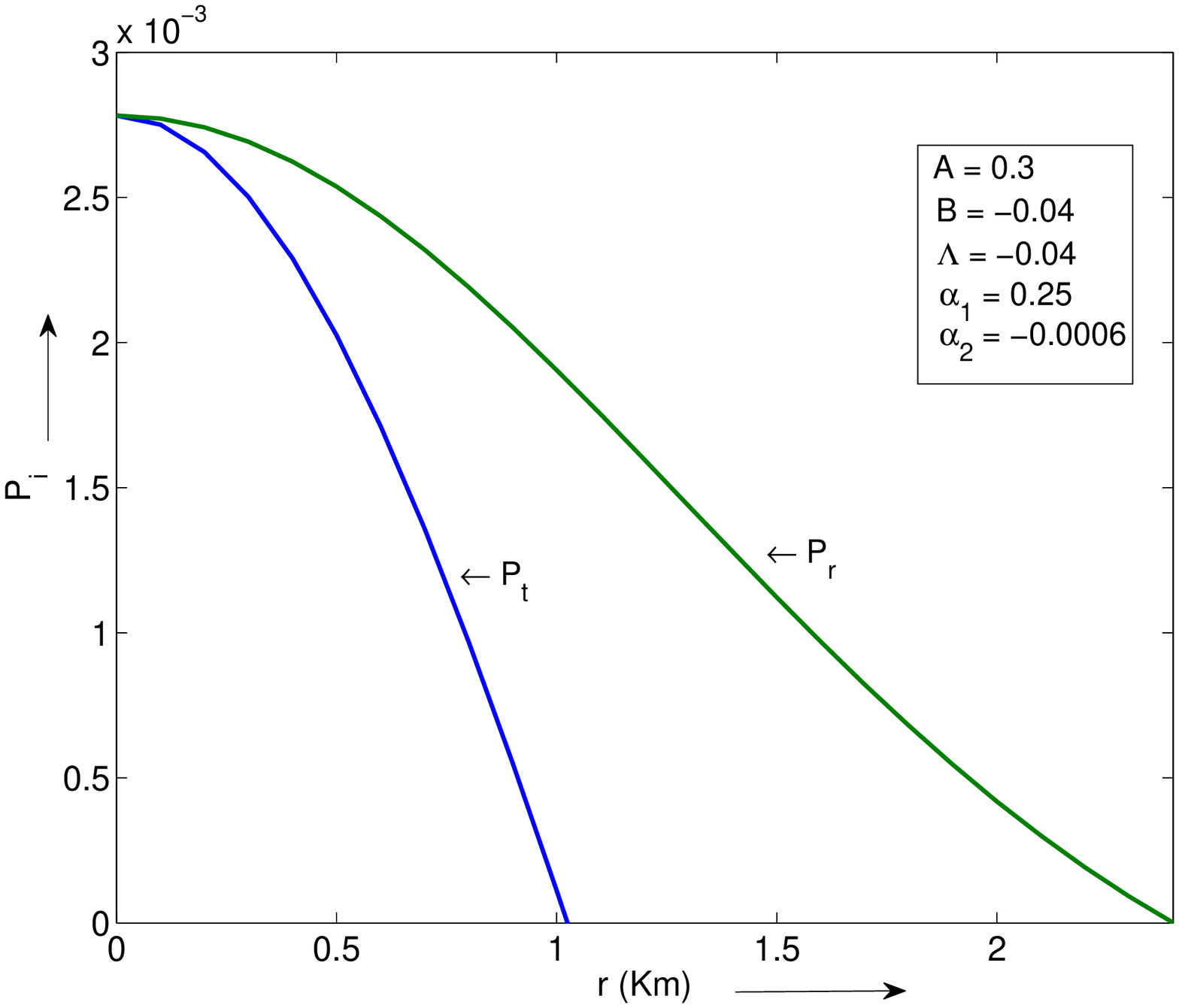}&
\includegraphics[width=9cm]{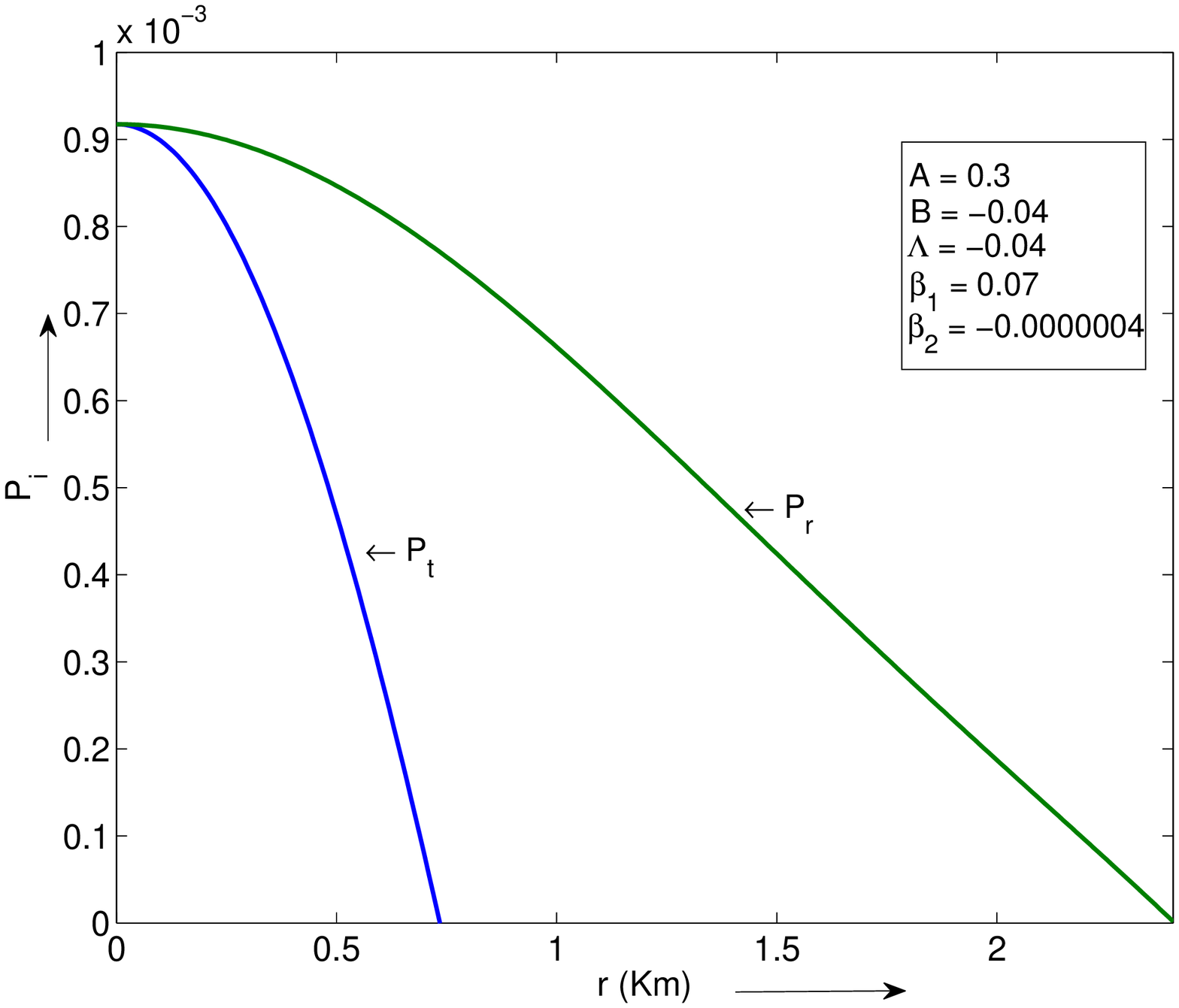} \\
\end{tabular}
\caption{ The radial and transverse pressure at the stellar interior for the linear (left) and non-linear (right) EOS.}
\end{figure*}

\subsection{Stability for linear EOS}
 In linear case, we have

\begin{eqnarray}
\frac{dp_r}{dr}&=& -\frac{\alpha_1}{4\pi}(A-B)Are^{-Ar^2} <~ 0, \label{eq35}\\
v^2_{sr} &=& \frac{dp_r}{d\rho} = \alpha_1,\label{eq36}\\
v^2_{st} &=& \frac{dp_t}{d\rho} = \left[-\frac{\left(\alpha_1(A-B)-B\right)\left(\alpha_1(A-B)-(A+B)\right)(1-Ar^2)}{A(A-B)}
-\frac{64\pi^2 \alpha_2^2}{A(A-B)}\left(1+Ar^2\right)e^{2Ar^2}
     \right.\nonumber\\
&& \left. -\frac{8\pi \alpha_2}{(A-B)}e^{Ar^2}-\frac{\left(\alpha_1(A-B)-B\right)\left(16\pi\alpha_2e^{Ar^2}-A\right)}{A(A-B)}-\frac{B}{A-B} \right].\label{eq37}
\end{eqnarray}

\subsection{Stability for non-linear EOS}
In non-linear case, we have

\begin{eqnarray}
\frac{dp_r}{dr}&=& -\beta_1 \left(\frac{A-B}{4\pi}\right)Are^{-Ar^2} +\beta_2\left(\frac{16\pi}{A-B}\right)Are^{Ar^2}<~0, \label{eq38}\\
v^2_{sr} &=& \frac{dp_r}{d\rho} = \beta_1-\beta_2\left(\frac{8\pi}{A-B}e^{Ar^2}\right)^{2},\label{eq39}\\
v^2_{st} &=& \frac{dp_t}{d\rho} = \left[-\left(\beta_1(A-B)-B\right)^{2}\left(\frac{1-Ar^2}{A(A-B)}\right)
+\left(\beta_1(A-B)-B\right)\left(\frac{2-Ar^2}{A-B}\right)-\frac{\beta_2}{2}\left(\frac{128\pi^2}{A-B}\right)
\left(\frac{4+3Ar^2}{A-B}\right)e^{2Ar^2}
     \right.\nonumber\\
&& \left. -\frac{B}{A-B}-\frac{\beta_2^{2}}{8}\left(\frac{128\pi^2}{A-B}\right)^{2}\left(\frac{2+3Ar^2}{A(A-B)}\right)e^{4Ar^2}
  -{\beta_2}\left(\beta_1(A-B)-B\right)\left(\frac{128\pi^2}{A-B}\right)
        \left(\frac{1+Ar^2}{A(A-B)}\right)e^{2Ar^2} \right].\label{eq40}
\end{eqnarray}

 Fig. (2), indicates that the pressure (radial and tangential) gradients are negative and
monotonically decreasing functions of r. The necessary and sufficient criterion for the
adiabatic speed of sound to be less than unity is satisfied by the suitable choice of the
parameters shown in Fig. (3), however, Caporaso and Brecher \cite{Caporaso} claimed that
$\frac{dp}{d\rho}$ does not represent the signal speed. Therefore, if the speed of sound
exceeds the speed of light, this does not necessary mean that the fluid is non-causal.

\begin{figure*}[thbp]
\begin{tabular}{rl}
\includegraphics[width=9cm]{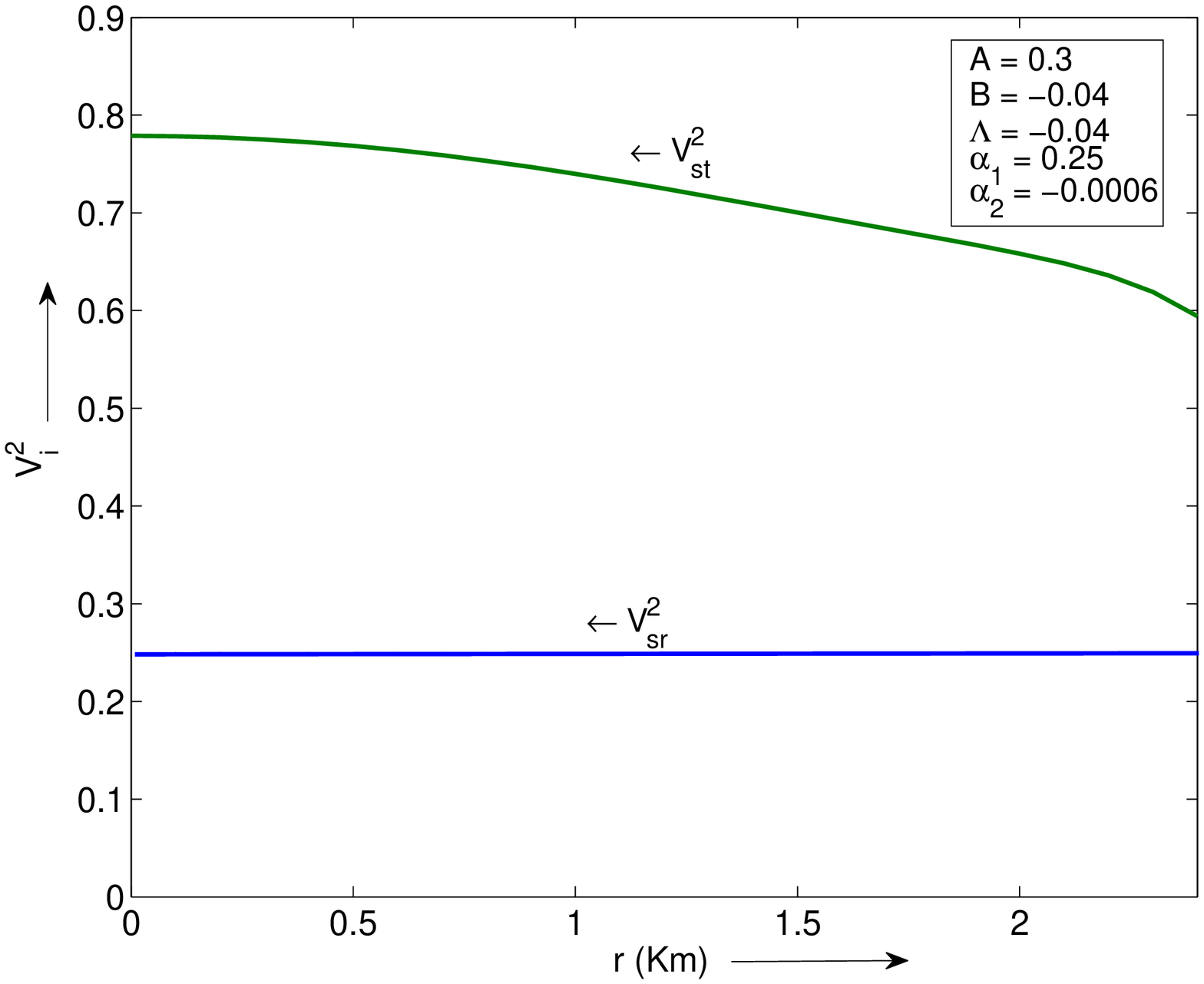}&
\includegraphics[width=9cm]{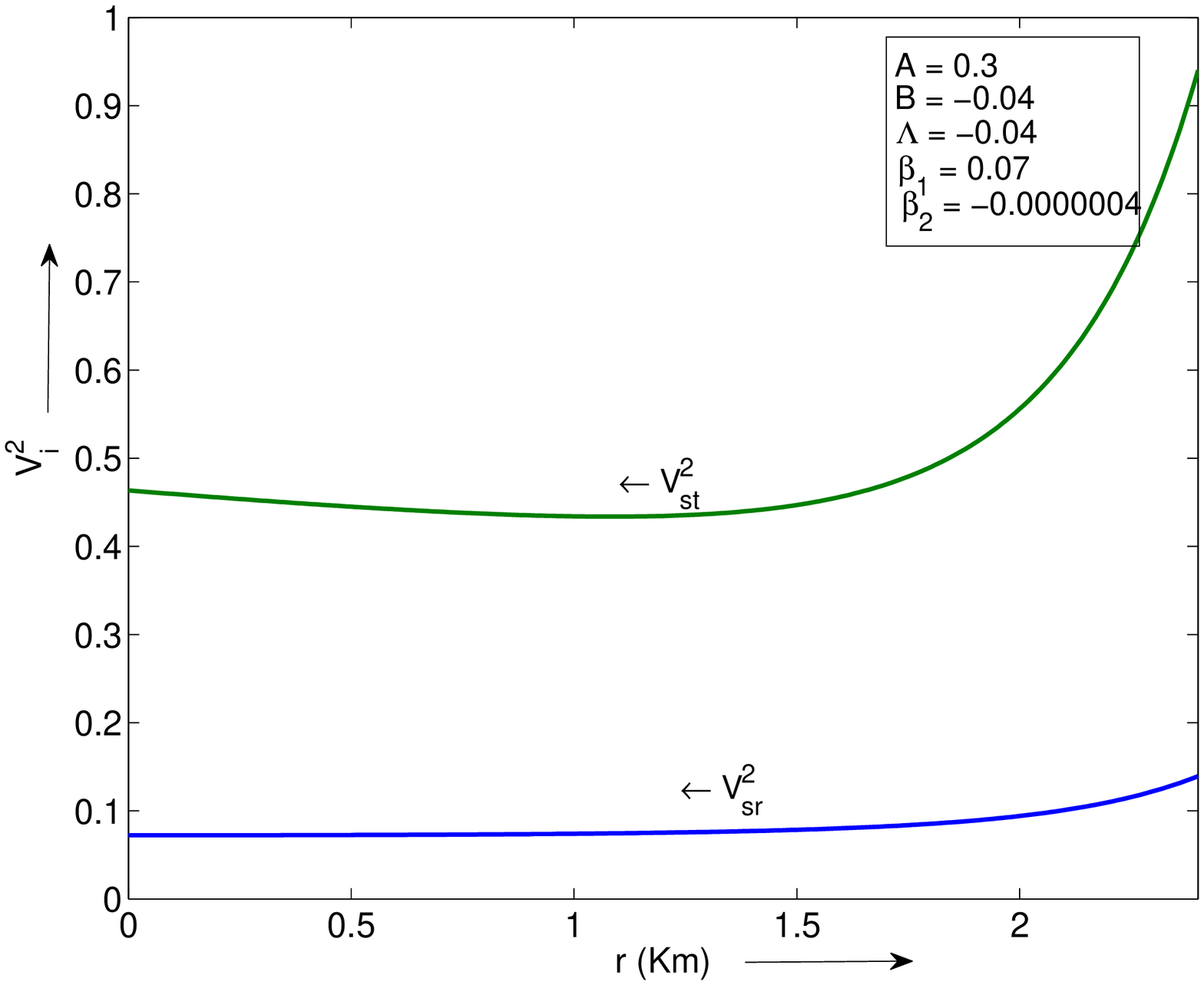}\\
\end{tabular}
\caption{The variation of radial sound speed $v^2_{sr}$ and tangential
sound speed $v^2_{st}$ sound speed at the stellar interior for the linear (left) and non-linear (right) EOS.}
\end{figure*}

\section{Energy Conditions}
The material composition of a physically reasonable charged fluid sphere has to obey
the null energy condition (NEC), weak energy condition (WEC)
and strong energy condition (SEC), at all points in the interior of a star
if the following inequalities hold simultaneously :
\begin{eqnarray}
\rho(r)+\frac{E^2}{8\pi} \geq  0 ,\label{eq41}\\
\rho(r)+p_r(r) \geq  0,\label{eq42}\\
\rho(r)+p_t(r)+\frac{E^2}{4\pi} \geq 0 ,\label{eq43}\\
\rho(r)+p_r+p_t(r)+\frac{(E^2-\Lambda)}{8\pi} \geq  0 .\label{eq44}
\end{eqnarray}
It is to be noted in Table 1, that the set of  constant parameters follow
the restrictions and satisfy all the energy conditions shown in
Fig. (4) for both linear and non-linear cases.

\begin{table}
\center
  \begin{tabular}{c|cccccc}\hline\hline
    EOS & A(+ ve) & B (- ve) & $\Lambda$ (- ve) & $M_0$ & Q & R(approx.)\\ \hline
          linear & 0.3 & 0.04 & 0.04 & 0.045 & 0.1 & 2.401\\  \hline
          non-linear & 0.3 & 0.04 & 0.04 & 0.045 & 0.1& 2.404 \\ \hline
  \end{tabular}
\caption{ Values of the constant parameters.}
\end{table}

\begin{figure*}[thbp]
\begin{tabular}{rl}
\includegraphics[width=9cm]{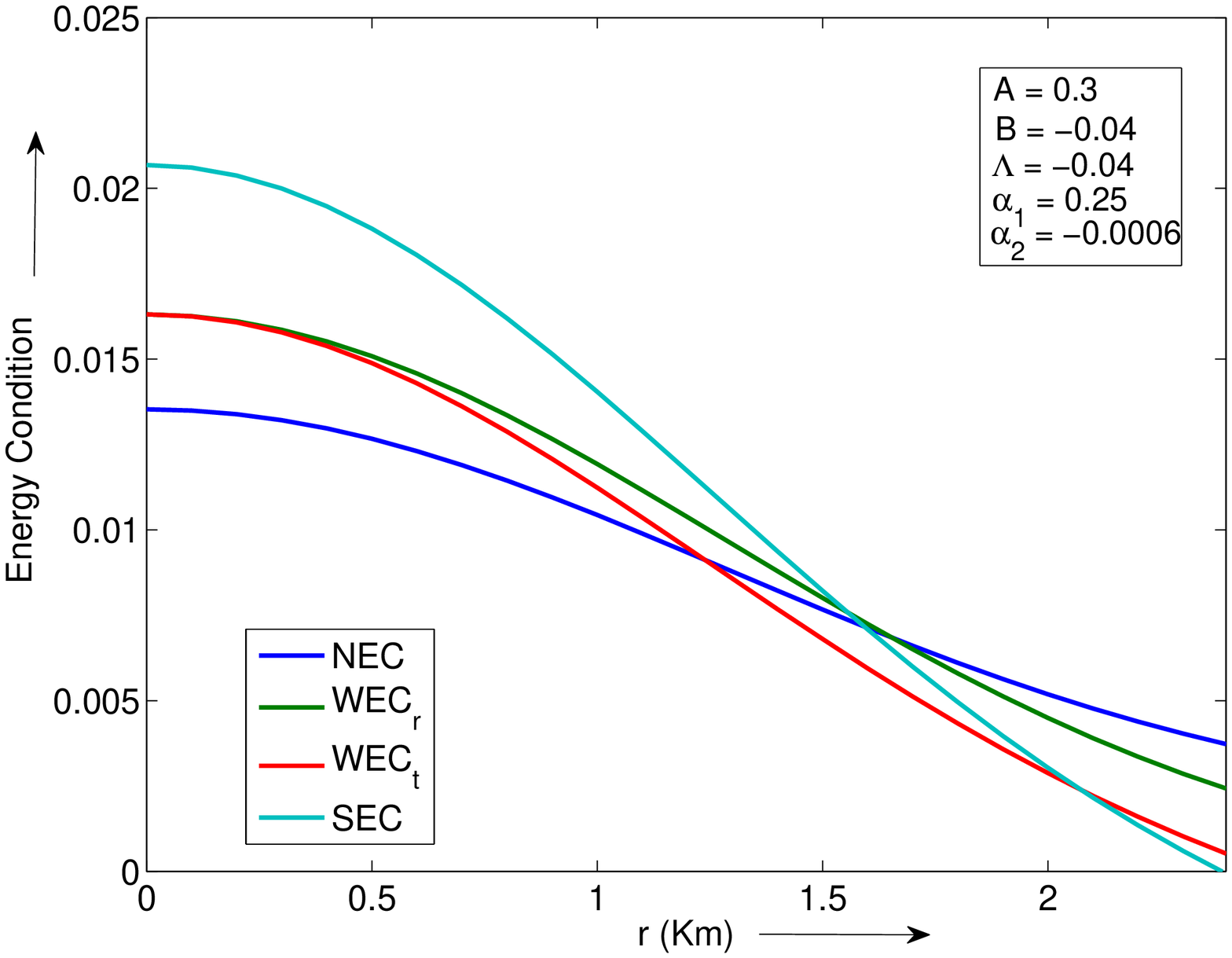}&
\includegraphics[width=9cm]{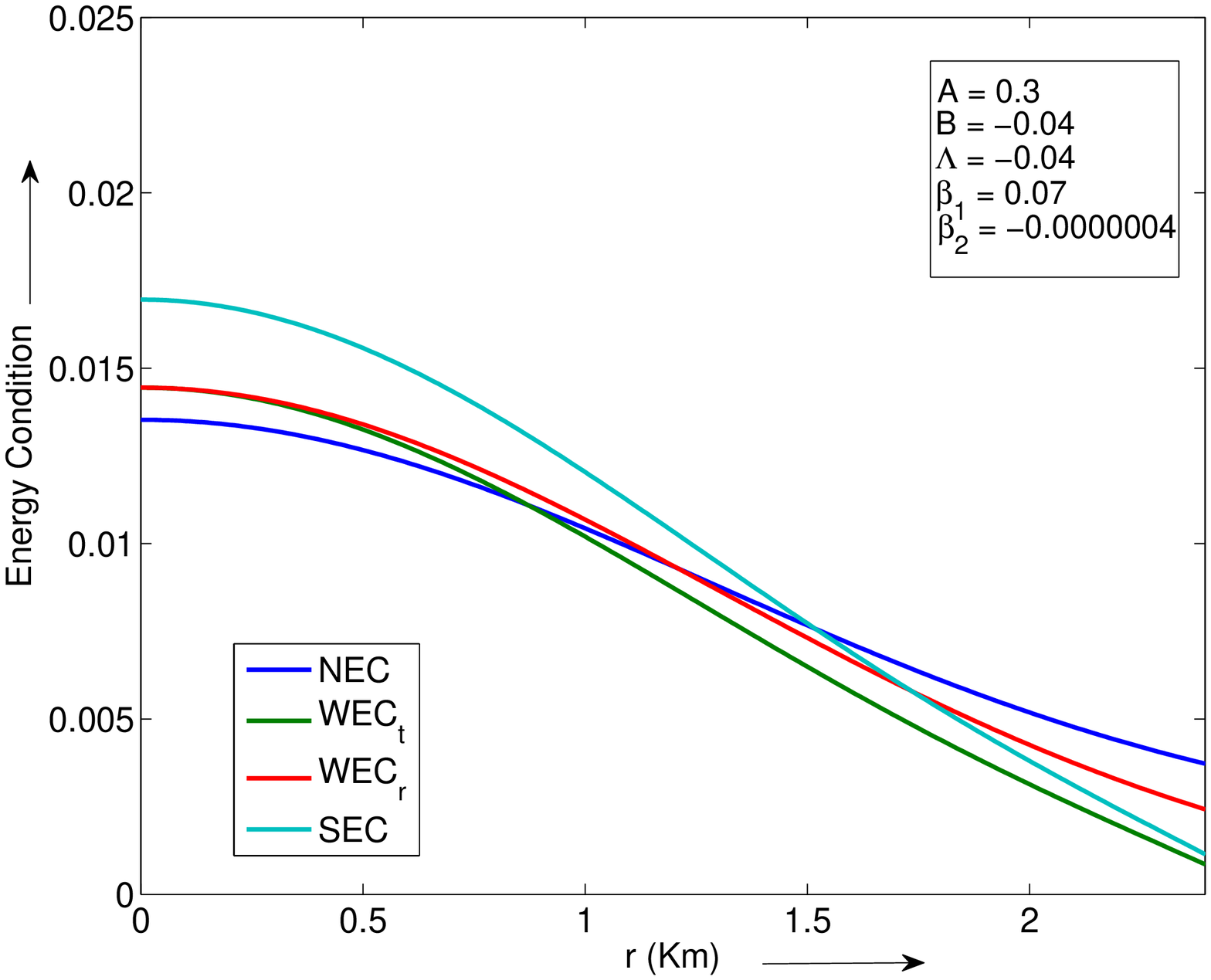}\\
\end{tabular}
\caption{ This figures shows that all energy conditions are satisfied at all points in the interior of the star solutions for both linear (left) and non-linear (right) EOS.  }
\end{figure*}

\subsection{Gravitational effect on mass-radius relation and compactness}
In this section, we study the maximum allowable mass-radius
ratio for our model and the upper limit on the mass may be written
as
\begin{equation}
8m(r) \equiv 1-e^{-Ar^2}-\Lambda r^2 \leq 1-e^{-AR^2}-\Lambda R^2.\label{eq45}
\end{equation}

In Fig. (5), we have plotted  $\frac{m(r)}{r}$ against $r$, which shows that
the ratio $\frac{m(r)}{r}$ is an increasing function of the radial
parameter. We observe that a constraint on the
maximum allowed mass-radius ratio in our case falls within the limit
of the (3+1)-dimensional isotropic fluid sphere i.e.,
$\left(\frac{m(r)}{r}\right)_{max} = 0.0548$  $<\frac{4}{9}$, as
obtained by Buchdahl \cite{Buchdahl}.

 From Eq. (9), the compactness of the stellar configuration is given
 by
\begin{equation}
\label{eq46} \mathcal{U}= \frac{ m(r)} {r}\Bigl\lvert_{r=R}=\frac{1}{8r}\left(1-\Lambda
r^2-e^{-Ar^2}\right),
\end{equation}
and the surface redshift $\mathcal{Z}_s$ corresponding to the above compactness ($\mathcal{U}$)
is obtain as

\begin{equation}
\label{eq47} \mathcal{Z}_s= ( 1-8 \mathcal{U})^{-\frac{1}{2}} - 1,
\end{equation}
where
\begin{equation}
\label{eq48} \mathcal{Z}_s=  \left[1-\frac{1}{r}\left(1-\Lambda
r^2-e^{-Ar^2}\right)\right]^{-\frac{1}{2}}-1.
\end{equation}

Thus, the maximum surface redshift $Z_s$ of our (2+1)-dimensional
stellar configuration with radius R= 2.4 turns out to be $\mathcal{Z}_s= 0.335$.
For a static perfect fluid sphere in (3+1) dimensional gravity the surface redshift is less
than Z = 2 \cite{Buchdahl}, however for anisotropic spheres this value may be larger \cite{Ivanov}.

\begin{figure*}[thbp]
\begin{tabular}{rl}
\includegraphics[width=9cm]{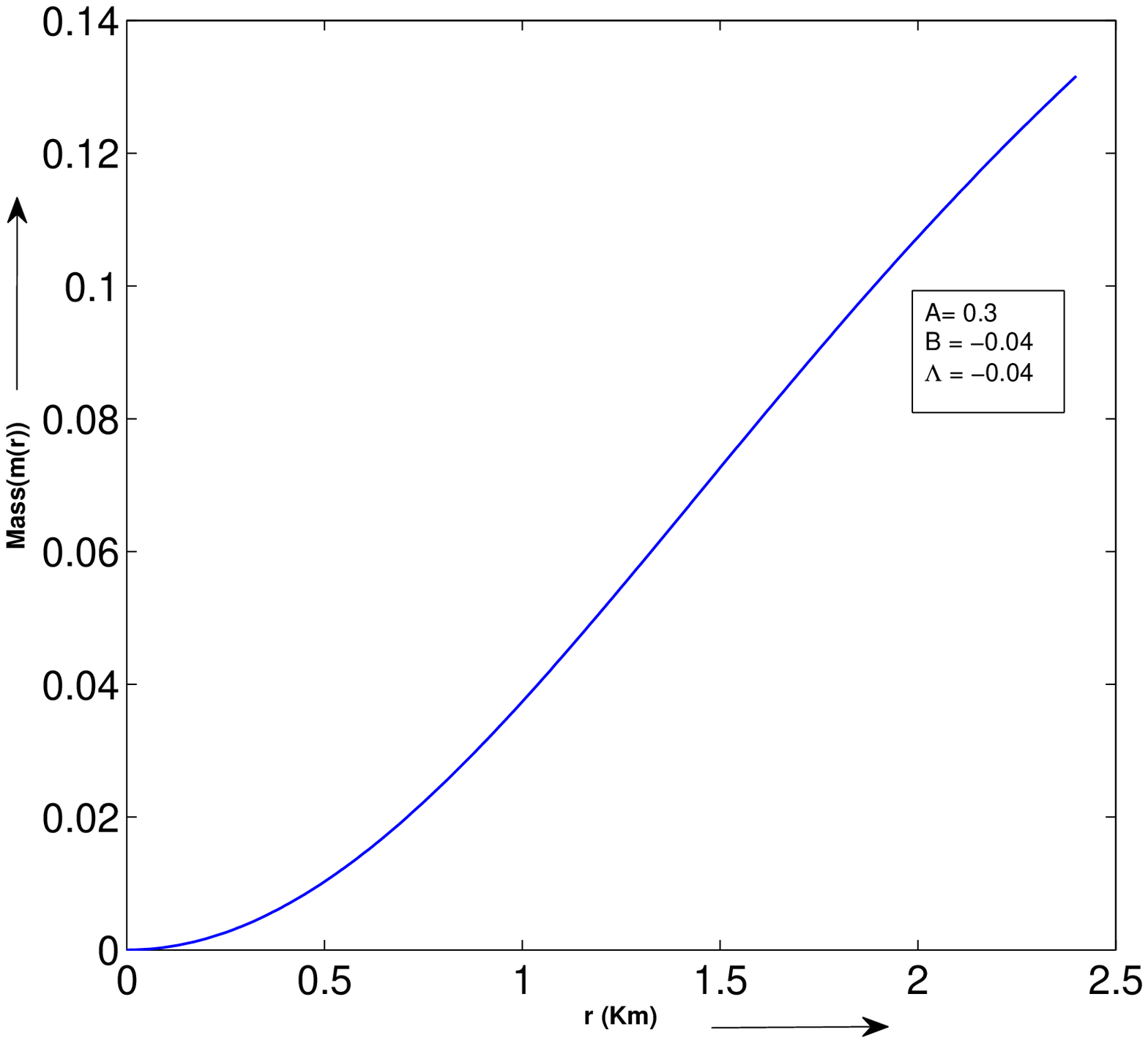}&
\includegraphics[width=9cm]{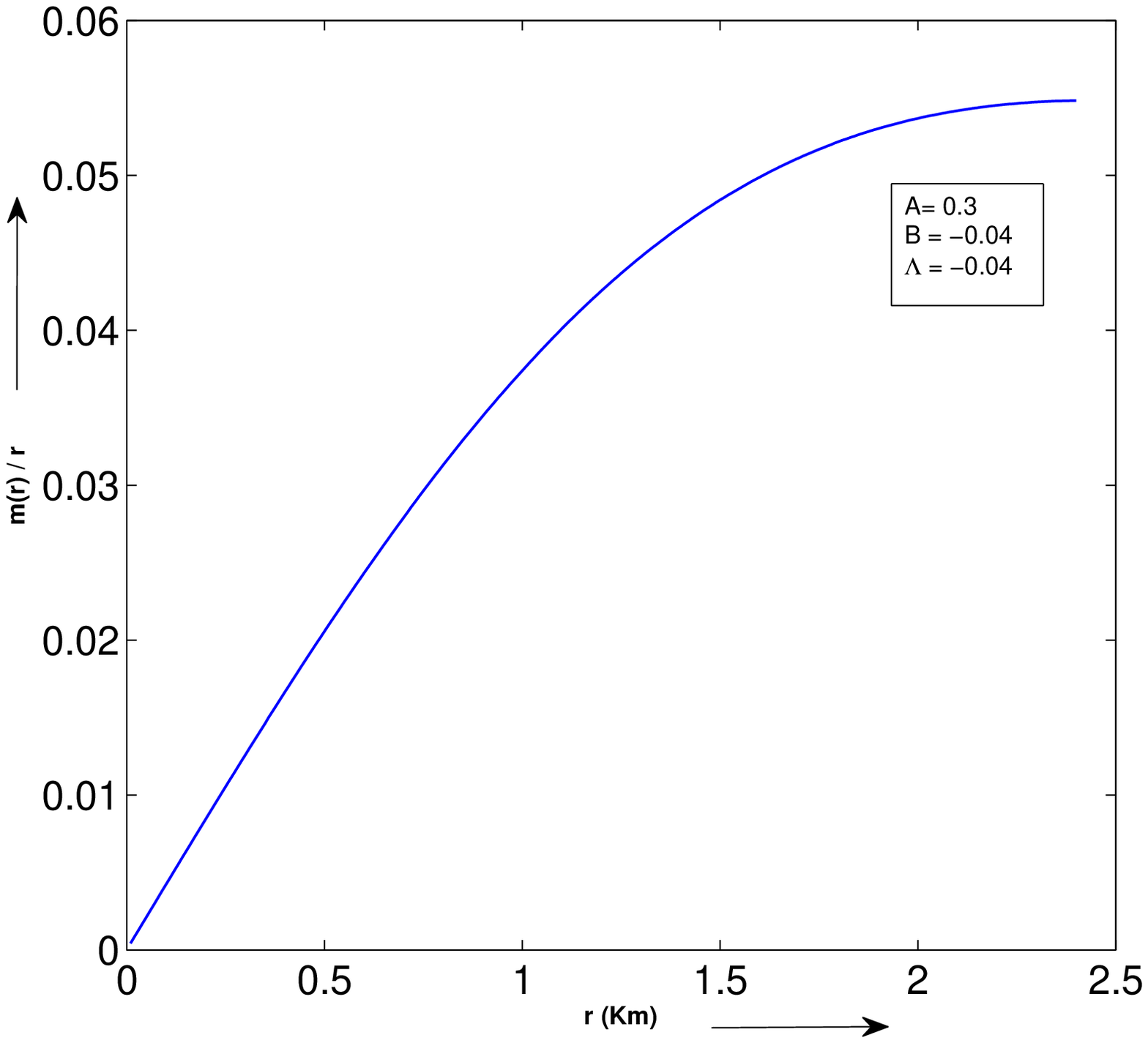}\\
\end{tabular}
\caption{ Variation of mass (left) and $\frac{m(r)}{r}$ (right) functions at stellar interior. }
\end{figure*}

\section{Equilibrium configuration}

In Section III, we have matched the interior solution to
the exterior charged BTZ metric with $p = \rho =0$, at a junction interface
$\Sigma $, with junction radius r = R. In our case, the junction surface is an one dimensional
ring of matter. One may also impose the boundary and regularity conditions,
i.e., the metric coefficient are continuous across the surface, and their derivatives may
not be continuous at the surface. In other words, the
affine connections may be discontinuous at the boundary
surface where the two boundary layers will both have a non-zero
stress-energy. The magnitude of this
stress-energy can be computed in terms of the second fundamental forms
 at the boundaries. So we adopt $\eta$, the Riemann normal
coordinates at the junctions is  positive in the manifold described
by exterior Charged BTZ space-time and negative in the
manifold described by the interior space-time with $x^\mu = ( \tau,\phi,\eta) $,
where $\tau$ represents the proper time on the shell. The second fundamental forms
are given by:

\begin{equation}K^{i\pm}_j =  \frac{1}{2} g^{ik}
\frac{\partial g_{kj}}{\partial \eta}   \Bigl\lvert_{\eta =\pm 0} =
\frac{1}{2}   \frac{\partial r}{\partial \eta} \Bigl\lvert_{r=a}
 ~ g^{ik} \frac{\partial g_{kj}}{\partial r}\Bigl\rvert_{r=a}.
\label{eq49}
\end{equation}

Since, $K_{ij}$ is not continuous at the junction, so the second fundamental forms with
 discontinuity are given by
\begin{equation}\mathcal{K} _{ij} =   K^+_{ij}-K^-_{ij}.
\label{eq50}
 \end{equation}
Now, from Lanczos equation in (2+1) dimensional space-time,
the Einstein equations lead to

\begin{equation}
S^i_j=-\frac{1}{8\pi}\left(K^i_j-\delta^i_jK^k_k\right).
\label{eq51}
 \end{equation}

 Considering circular symmetry,  $K^i_j$ becomes
\begin{equation}
K^i_j =
\left( {\begin{array}{cc}
 K^\tau_\tau & 0  \\
 0 & K^\phi_\phi \\
 \end{array} } \right),
 \label{52}
\end{equation}
then one can obtain the surface stress energy tensor
$S^i_j $ =diag$ (-\sigma, -v)$
in terms of the energy density $\sigma$ and is surface pressure v,
then Lanczos equation in (2+1)-dimensional space-time \cite{Banerjee, A. Banerjee}, becomes

\begin{eqnarray}
\sigma &=&  -\frac{1}{8\pi}  \kappa _\phi^\phi,\\
\label{eq53} v &=&  -\frac{1}{8\pi}  \kappa _\tau^\tau.
\label{eq54}
 \end{eqnarray}
Using the Eq. (52) in Eq.~(\ref{eq51}) we obtain
\begin{equation}
\sigma =  -\frac{1}{8\pi R} \left[ \sqrt{M_0-\Lambda R^2- Q^2\ln R}+e^{-\frac{AR^2}{2}} \right],\label{eq55}
 \end{equation}
corresponding to the linear EOS
 \begin{equation}
 v =  -\frac{1}{8\pi}  \left[ \frac{\Lambda R+\frac{Q^2}{2R}}{\sqrt{M_0-\Lambda R^2- Q^2\ln R}} +\{2\left(\alpha_1(A-B)-B\right)Re^{-\frac{AR^2}{2}}+16\pi \alpha_2 Re^{\frac{AR^2}{2}}\}\right],\label{eq56}
\end{equation}
or, for non-linear EOS
 \begin{equation}
  v =  -\frac{1}{8\pi}  \left[ \frac{\Lambda R+\frac{Q^2}{2R}}{\sqrt{M_0-\Lambda R^2- Q^2\ln R}} +\{2\left(\beta_1(A-B)-B\right)Re^{-\frac{AR^2}{2}}+4\beta_2\left(\frac{32\pi^2}{A-B}\right) Re^{\frac{3AR^2}{2}}\}\right],\label{eq57}
\end{equation}
where, we have set r = R.  Lobo in his work \cite{Lobo} shown that we can introduction of a
thin shell on the junction hypersurface and it is possible to match the interior
and exterior space-time without a thin shell.
As we match second fundamental forms associated with the two
sides of the shell, a crucial question arises in the
form of the star's stability against collapse. Therefore, we
must have a thin ring of matter component with above
stresses so that the outer boundary exerts outward force
to balance the inward pull of BTZ exterior. The energy
density is negative in this junction ring which is similar
to the (3 + 1) dimensional case \cite{Visser}.

\section{Conclusion}

In this work we have found a new type of charged star solution treating the
matter distribution as anisotropic in nature admitting linear or
nonlinear equation of state and for a specific field intensity.
The starting point of this work is to  assume a specific form of mass function m(r)
and using this we found the energy density at the stellar interior. We match the star solution to
the exterior charged BTZ solution and found the values of physically reasonable constant
terms. All the solutions, which are obtained in both linear and nonlinear EOS are regular at the
center and well behaved in the stellar interior except at the boundary where we propose
a thin ring of matter content with negative energy density so as to prevent collapsing.

 We observe that the energy density ($\rho$) and the
pressures (radial and transverse) are positive and monotonically decreasing functions in the
stellar interior, where the electric field intensity ($E^2$) and mass (m(r)) functions
are positive and monotonically increasing within the radius of the star R=2.4 as shown
in figures corresponding to the particular choices of the constant parameters given in Table 1.
Consequently the solutions which we obtained in this paper has a simple analytical
form which can be used to study a full collapsing model of a (2 + 1)-dimensional
charged relativistic fluid sphere which is beyond the scope of this analysis.

\end{document}